\documentclass[aps,onecolumn,showpacs,groupedaddress, nofootinbib, 11pt]{revtex4}  
\usepackage{graphicx}
\usepackage{epstopdf}
\usepackage{amsmath}
\usepackage{amsfonts}
\usepackage{amssymb}
\usepackage{appendix}
\usepackage{mathtools}
\usepackage{comment}
\usepackage{bbold}
\usepackage{color}
\usepackage{slashed}
\usepackage{subfigure}
\usepackage{setspace}
\usepackage{footnote}
\usepackage[activate={true,nocompatibility},final,tracking=true,kerning=true,spacing=true,factor=1100,stretch=10,shrink=10]{microtype}

\begin{document}

\singlespacing

\hfill NUHEP-TH/14-09

\title{Heavy Neutrinos and the Kinematics of Tau Decays} 

\author{Andrew Kobach} 
\author{Sean Dobbs} 
\affiliation{Northwestern University, Department of Physics \& Astronomy, 2145 Sheridan Road, Evanston, IL 60208, USA}

\date{\today}

\begin{abstract} 
Searches for heavy neutrinos often rely on the possibility that the heavy neutrinos will decay to detectable particles. Interpreting the results of such searches requires a particular model for the heavy-neutrino decay.  We present a method for placing limits on the probability that a tau can couple to a heavy neutrino, $|U_{\tau4}|^2$, using only the kinematics of semi-leptonic tau decays, instead of a specific model.  Our study suggests that $B$ factories with large datasets, such a Belle and BaBar, may be able to place stringent limits on $|U_{\tau 4}|^2$ as low as $\mathcal{O}(10^{-7} - 10^{-3})$ when 100 MeV $\lesssim m_4 \lesssim $ 1.2 GeV, utilizing minimal assumptions regarding the decay modes of heavy neutrinos.  
\end{abstract}

\pacs{13.35.Dx, 14.60.St}
\maketitle

\section{Introduction}

The explanation of neutrino masses requires degrees of freedom beyond those currently available in the standard model (SM).  A popular option is to augment the SM with new ``neutrinos'' whose masses can, in principle, exist anywhere between the eV and GUT scales.  This generic possibility offers the potential to address a broad range of open puzzles in particle physics, well beyond neutrino masses (for an extensive review, see Ref.~\cite{Drewes:2013gca} and references found therein).

In this work, we consider that heavy neutrinos can interact with the tau via charged-current weak interactions.  For simplicity, we take there to be only one such heavy neutrino, $\nu_4$.  Here, we let the probability that the tau interacts with $\nu_4$ to be $|U_{\tau4}|^2$, and the probability that the tau interacts with the known ``light'' neutrinos ($\nu_1,$ $\nu_2,$ $\nu_3$) to be $1-|U_{\tau4}|^2$.

Here, we summarize the relatively few sources of constraints on the value of $|U_{\tau4}|^2$, all of which assume $\nu_4$ can interact with SM particles via the weak interactions.  
Limits are estimated by NOMAD~\cite{Astier:2001ck} and CHARM~\cite{Orloff:2002de} experiments, which have detectors located downstream from a beam of high-energy protons incident on a fixed target.  Under the assumption that $\nu_4$ can decay primarily via neutral-current weak interactions, these two experiments search for the signatures associated with $\nu_4$ decay within the detectors' fiducial region. 
The DELPHI experiment~\cite{Abreu:1996pa} at LEP estimates limits on the value of $|U_{\tau4}|^2$ by searching for signatures of a (mostly) sterile $\nu_4$ that decays to ``visible'' SM particles in $e^+ e^- \rightarrow Z \rightarrow \nu \nu_4$ events.
Lastly, the authors of Ref.~\cite{Helo:2011yg} use measurements of tau and meson branching ratios to estimate limits on $|U_{\tau4}|^2$, assuming that the mass and lifetime of the tau are known to infinite precision.
All of the aforementioned constraints can be seen in Fig.~\ref{limitplot}.  Taken together, these studies estimate that the value of $|U_{\tau4}|^2 < \mathcal{O}(10^{-5} - 10^{-3} )$ for 50 MeV $\lesssim m_4 \lesssim$ 60 GeV, where $m_4$ is the mass of $\nu_4$.

These analyses all utilize assumptions regarding the possible branching ratios of $\nu_4$.  It is possible, however, that one can search for the presence of a heavy neutrino without relying on a specific model that dictates its lifetime and decay modes. 
If the tau decays semi-leptonically into a neutrino and a hadronic system, $\tau^-\rightarrow \nu +h^-$ ($\nu$ is a mass eigenstate), then the possible energy and momentum of $h^-$, i.e., its kinematic phase space, itself can contain information whether it ``recoiled'' against a heavy neutrino.\footnote{Similar in spirit are analyses that place limits on the ``mass of the tau neutrino,'' e.g., ALEPH~\cite{Barate:1997zg} and CLEO~\cite{Athanas:1999xf}.  The interpretation of the results from these experiments is nontrivial, since we now know that the ``tau neutrino'' is not a mass eigenstate.  This is discussed further in Section~\ref{kinematics}.}   The kinematic phase space of $h^-$ could be the superposition of two possibilities:~the phase space associated with a heavy neutrino, weighted by $|U_{\tau4}|^2$, and the phase space associated with effectively-massless neutrinos, weighted by $(1-|U_{\tau4}|^2)$. 
Searching for heavy neutrinos using only the information contained in $h^-$ can be, to a good approximation, insensitive to the details of $\nu_4$ decay and whether it is Dirac or Majorana.

This method to search for heavy neutrinos using the hadronic system in tau decays requires high statistics and good momentum resolution, both of which are possible at $B$ factories.  We investigate potentially-achievable limits on $|U_{\tau4}|^2$ by creating simulated pseudo-data of the process $e^+e^- \rightarrow \tau^+\tau^-$ at $\sqrt{s}=11$ GeV, where one of the taus decays as $\tau^-\rightarrow \nu + h^-$, where $\nu$ is any of the four neutrino mass eigenstates and $h^-$ is comprised of $\pi^-\pi^+\pi^-$.
We find that experiments with large data samples, such as Belle and BaBar, could place competitive limits, e.g., $|U_{\tau4}|^2 <\mathcal{O}(10^{-7}-10^{-3})$, when $100\text{ MeV} \lesssim m_4 \lesssim 1.2$ GeV.  Such a result would depend on minimal theoretical assumptions.

Our work is outlined as follows.  In Section~\ref{kinematics}, we discuss the kinematics of semi-leptonic tau decays and show how a final-state hadronic system can can contain information regarding whether it ``recoiled'' against a heavy neutrino.  In Section~\ref{analysis}, we discuss our pseudo-data simulation at a $B$ factory and estimate a range of limits on $|U_{\tau4}|^2$ that experiments may be able to achieve.  In Section~\ref{conclusion}, we discuss results and offer concluding thoughts.

\section{Kinematics of Tau Decays}
\label{kinematics}

\begin{figure}[htbp]
\begin{center}
\includegraphics[width=0.7\textwidth]{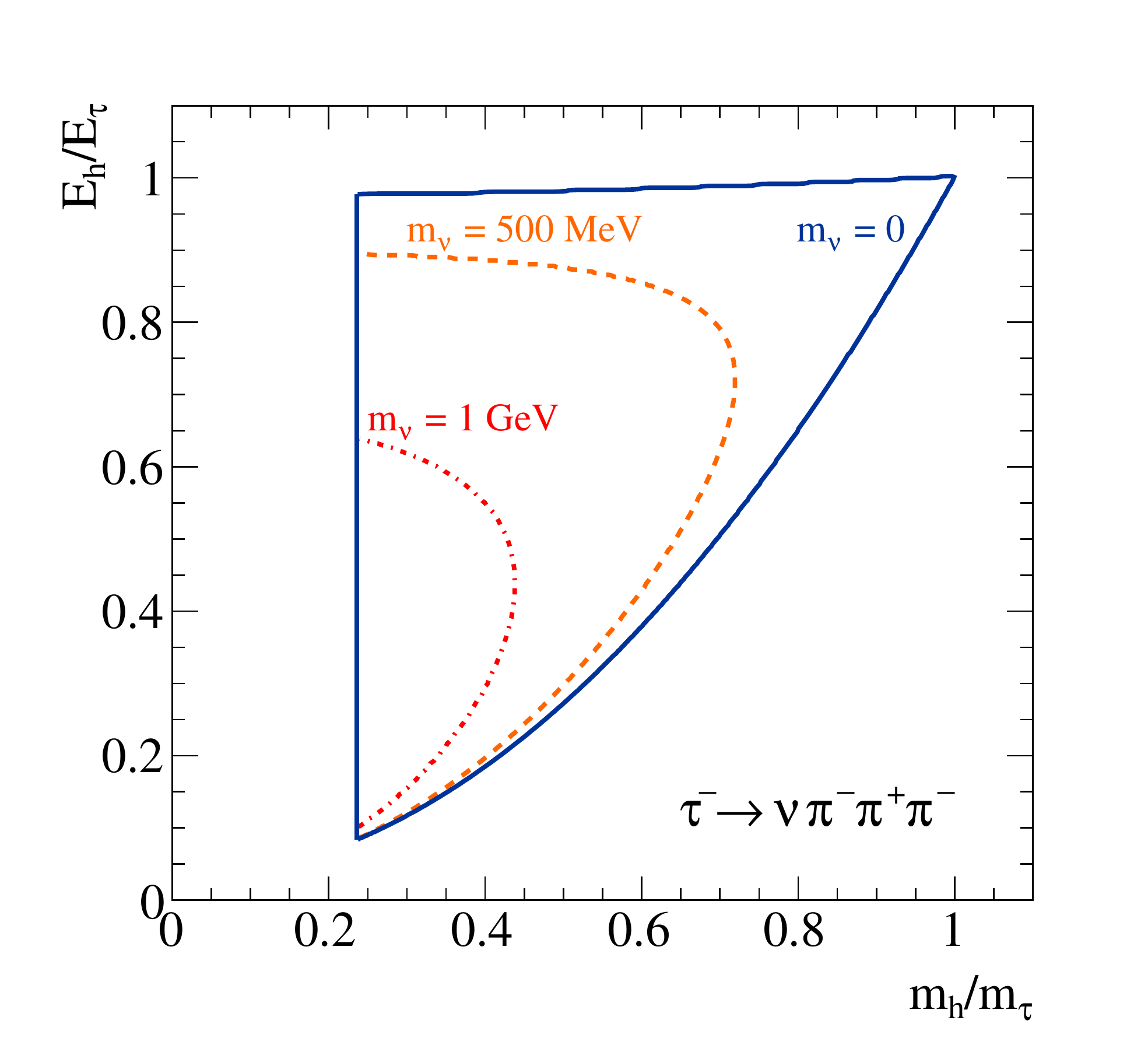}
\caption{The available kinematic phase space for $E_h/E_\tau$ and $m_h/m_\tau$ in $\tau^-\rightarrow \nu \pi^-\pi^+\pi^-$, where the invariant mass of the hadronic system is in the range $3m_{\pi^\pm} < m_h < m_\tau - m_\nu$.  For a given value of $m_\nu$, the range of $E_h$ is given by Eqs.~(\ref{ehmax}) and~(\ref{ehmin}). We show the available phase space when the mass of the final-state neutrino is zero (dark blue), 500 MeV (orange), and 1 GeV (red). }
\label{psplot}
\end{center}
\end{figure}

Here, we analyze how a single heavy neutrino can alter the kinematics of semi-leptonic tau decays, $\tau^-\rightarrow \nu + h^-$.  If the hadronic system is comprised of multiple particles, the range of possible values for its invariant mass ($m_h$) and energy ($E_h$) changes as a function of $m_\nu$.\footnote{While the values of $m_h$ and $E_h$ are correlated, more information can be extracted by considering both variables instead of one or the other.}
If the hadronic system $h^-$ hadronizes into charged pions or kaons, then reconstructing the values of $m_h$ and $E_h$ is possible at high precision.  
For a given value of $m_h$, the range of $E_h$ is given by
\begin{eqnarray}
\label{ehmax}
E_h^\text{max} &=& E_\tau - \sqrt{m_\nu^2 + q^2_+}, \\
E_h^\text{min} &=& E_\tau - \sqrt{m_\nu^2 + q^2_-}, 
\label{ehmin}
\end{eqnarray}
where
\begin{equation}
q_\pm \equiv \frac{m_\tau}{2} \left( \frac{m_h^2-m_\tau^2-m_\nu^2}{m_\tau^2} \right) \sqrt{\frac{E_\tau^2}{m_\tau^2} -1 }~ \pm ~\frac{E_\tau}{2} \sqrt{ \left( 1-\frac{(m_h+m_\nu)^2}{m_\tau^2}\right) \left( 1-\frac{(m_h-m_\nu)^2}{m_\tau^2}\right) }.
\end{equation}
As an illustration, consider the process $e^+e^-\rightarrow \tau^+\tau^-$, where one of the taus decays like $\tau^-\rightarrow \nu \pi^-\pi^+\pi^-$. Here, $E_\tau = E_\text{beam}/2$ in the limit of no initial-state radiation. The value of $m_h$ can exist, in principle, in the range $3m_{\pi^\pm} < m_h < m_\tau-m_\nu$. The range of $E_h$ is given by Eqs.~(\ref{ehmax}) and~(\ref{ehmin}).  The available phase space for $E_h/E_\tau$ and $m_h/m_\tau$ is shown in Fig.~\ref{psplot}, when $m_\nu= 0$, 500 MeV, and 1 GeV.

The  phase space of the hadronic system can be described as a linear superposition of two distinct distributions:~one for effectively-massless neutrinos, weighted by $(1-|U_{\tau4}|^2)$, and one for a heavy neutrino with mass $m_4$, weighted by $|U_{\tau4}|^2$, i.e., 
\begin{equation}
\label{weighting}
\frac{d\Gamma_\text{tot}(\tau^-\rightarrow \nu h^-)}{dm_hdE_h} = \left(1-|U_{\tau4}|^2\right) \frac{d\Gamma(\tau^-\rightarrow \nu h^-)}{dm_hdE_h} \Big|_{m_\nu = 0} + |U_{\tau4}|^2 \frac{d\Gamma(\tau^-\rightarrow \nu h^-)}{dm_hdE_h} \Big|_{m_\nu = m_4}.
\end{equation}
The presence of a heavy neutrino introduces a curved, crescent-shaped endpoint structure in the $E_h-m_h$ phase space for the largest values of $m_h$.  Therefore, analyzing the measured shape of the $E_h - m_h$ phase space can allow for the possibility of constraining the value of $\left|U_{\tau4}\right|^2$.

The kinematics of tau decays, as discussed here, have been studied at length by experiments like ALEPH~\cite{Barate:1997zg} and CLEO~\cite{Athanas:1999xf}.  These experiments use 3- and 5-prong tau decays to place limits on the ``mass of the tau neutrino,'' under the assumption that the tau interacts with only a single, massive neutrino.  Understanding of the neutrino sector has advanced significantly since the time of these analyses; we believe that the tau interacts with $\nu_1$, $\nu_2$, and $\nu_3$, all of which are effectively massless.  Consequently, the results from ALEPH and CLEO are somewhat ambiguous to interpret in a modern context.  We repurpose these kinematic methods to study the possibility of constraining the existence of a heavy neutrino in tau decays.

\section{Pseudo-data Analysis}
\label{analysis}

We estimate limits on $|U_{\tau4}|^2$, as a function of $m_4$, by creating a pseudo-data sample of $e^+e^-\rightarrow \tau^+\tau^-$, with $\sqrt{s} = 11$ GeV, made using Monte Carlo (MC) simulations.  
At least one of the taus is required to decay as $\tau^-\rightarrow \nu \pi^-\pi^+\pi^-$.   This decay channel is chosen because of the excellent momentum resolution and a large branching fraction.\footnote{Another possibility is $\tau^-\rightarrow \nu 3\pi^-2\pi^+$ events, but because of the suppressed phase space due to the multiplicity of final-state particles, decays of this type may only aid in placing limits if $10\text{ MeV} \lesssim m_4 \lesssim 50\text{ MeV}$. }  This hadronic system permits one to study $0 < m_4 < m_\tau-3m_{\pi^\pm}$, however, this range is not fully experimentally accessible, because the hadronic phase space is not sufficiently different from the one for $m_4=0$ if $m_4\lesssim 10$ MeV, and the event rate is too small when $m_4 \gtrsim 1.2$ GeV.  Thus, we are only able to estimate meaningful limits if $10\text{ MeV} \lesssim m_4 \lesssim 1.2$ GeV.

We use the MC generator {\sc tauola}~\cite{Was:2004dg} with {\sc kk2f}~\cite{Jadach:1999vf} and {\sc photos}~\cite{Barberio:1993qi} to simulate $\sim$10M $\tau^-\rightarrow \nu \pi^-\pi^+\pi^-$ decays in a typical $B$-factory environment.   This sample size is chosen to correspond roughly with current data samples available at Belle~\cite{Lee:2010tc, Bevan:2014iga}.  This decay is dominated by $\tau\rightarrow \nu a_1$, and we set the mass and width of the $a^-_1(1260)$ to be 1250 MeV and 600 MeV, respectively.  This choice of model agrees well with experimental data~\cite{Lee:2010tc}.  We smear the momentum of the final-state pions to have a typical momentum resolution for $B$ factories, $\sigma/p = 0.1\% (p/\text{GeV}) \oplus 0.5\%$~\cite{Lee:2010tc}.  The signal efficiency is expected to be fairly flat as a function of $m_h$~\cite{Lee:2010tc}, thus we do not consider its effects.  A typical sample of reconstructed $\tau^-\rightarrow \nu \pi^- \pi^+ \pi^-$ events contain $\sim$10\% background contamination from processes like $\tau^-\rightarrow \nu \pi^- \pi^+ \pi^-\pi^0$, etc.~\cite{Lee:2010tc}.  The shape of the background contribution varies slowly and smoothly, and can be subtracted without introducing significant systematic uncertainties in the shape of the measured signal distribution~\cite{Lee:2010tc}, thus we ignore this effect for the purposes of our analysis.

\begin{figure}[htbp]
\begin{center}
\includegraphics[width=0.8\textwidth]{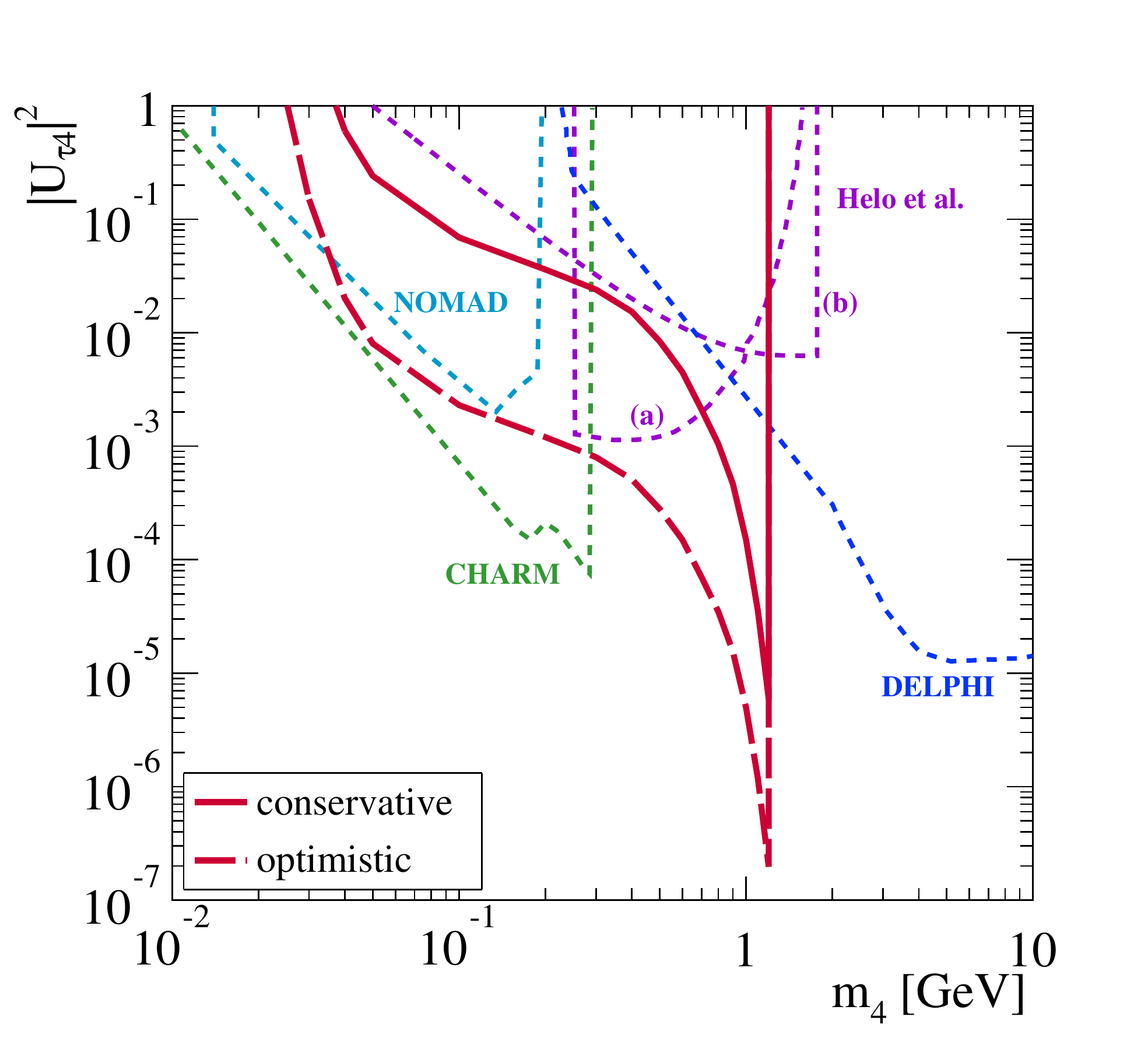}
\caption{ The red lines are the possible 95\% CL limits that $B$ factories may be able to achieve using an analysis of the kinematics of $\sim10$M $\tau^-\rightarrow \nu \pi^- \pi^+ \pi^-$ decays.  We note that these results depicted by these red lines are estimated using only MC simulations and are only an illustration of what could be possible at a $B$ factory.  Since the effects of systematic uncertainties are non-trivial, we estimate conservative and optimistic (meaning the systematic uncertainties are negligible) limits, as discussed in Section~\ref{analysis}.  The 90\% CL limits from the NOMAD~\cite{Astier:2001ck} (light blue) and CHARM~\cite{Orloff:2002de} (green) experiments are also shown, both of which assume that $\nu_4$ primarily decays via neutral-current interactions.  The results from the DELPHI (95\% CL)~\cite{Abreu:1996pa} experiment are shown in blue.   The results estimated in Ref.~\cite{Helo:2011yg}, utilizing measurements of tau and meson branching fractions, are shown in purple.  The line labeled (a) corresponds to the limits from interpreting the null results of ``visible'' $\nu_4$ decays within 10m, assuming the mass and lifetime of the tau are known to infinite precision. The line labeled (b) corresponds to limits estimated using the uncertainties associated with the purely leptonic branching ratios of the tau, assuming $|U_{e4}|^2 = |U_{\mu4}|^2=0$, that $\nu_4$ does not decay to ``visible'' particles within a typical detector environment, and that when $m_4 < m_\tau$, the data selection is not strongly affected by the presence of a heavy neutrino in the final-state. This latter limit, as shown here, is different than the one appearing in Ref.~\cite{Helo:2011yg}, since we require probability conservation, and we marginalize over the uncertainties associated with the mass and lifetime of the tau, as discussed in Appendix~\ref{fixedlimits}.   }
\label{limitplot}
\end{center}
\end{figure}

We make templates for a given value of $m_\nu$ by filling 2D histograms, as a function $m_h$ and $E_h$, with $\sim$500M $\tau^-\rightarrow \nu \pi^-\pi^+\pi^-$ events.
We weight a $m_\nu=0$ template by $(1-|U_{\tau4}|^2)$ and a $m_\nu=m_4$ template  by $|U_{\tau4}|^2$, summing the two, as expressed in Eq.~(\ref{weighting}).  With these, we use an unbinned log-likelihood function to compare to the pseudo-data  and estimate limits on $|U_{\tau4}|^2$, for a given value of $m_4$. 
The value of $2\Delta\ln\mathcal{L}$ is varied about its extremum by 3.84, and we take the corresponding value of $|U_{\tau4}|^2$ to be the 95\% CL.  
To test the bias due to the binning size, we increase and decrease the number of bins by a factor of two, and find that the results change negligibly.  These results are shown by the dashed-red line in Fig.~\ref{limitplot}.\footnote{If we used different model parameters of the $\pi^-\pi^+\pi^-$ hadronic current, then the results could be different than the ones presented here, though we do not expect that this would give rise to qualitative changes.}

Performing this study with true experimental data must address at least two important challenges.  First, one must select $e^+e^-\rightarrow \tau^+ \tau^-$ events with high efficiency and low background rate.  Second, the data selection must be inclusive enough to not veto events where the heavy neutrino could have decayed to ``visible'' particles within the detector.  For example, experiments could select events with at least one charged lepton ($e$ or $\mu$), missing energy, a $\pi^+\pi^-\pi^\pm$ system with tracks pointing back to the same position in space, and not vetoing on the presence of other particles in the event.  We presume that the templates and pseudo-data created for the purposes of our analysis do, to a good approximation, correspond to the results of an ``inclusive'' data selection at a $B$ factory.

Our limits presented thus far ignore the systematic uncertainties associated with the theoretical prediction of the qualities of the $\pi^-\pi^+\pi^-$ system.  We investigate these effects by creating new templates, increasing and decreasing the mass and the width of $a^-_1(1260)$  by $\approx 5\%$ and $\approx 15\%$, respectively, and redoing the analysis with the original pseudo-data sample.  The limits change significantly by varying these parameters;  the mass of the resonance has a particularly strong effect on the limits.  These variations are quite large compared to what would be typical for a true data analysis.   Because performing a detailed study of these systematic effects is highly non-trivial, we estimate a conservative limit on the value of $|U_{\tau4}|^2$, depicted by the red line in Fig.~\ref{limitplot}, which is a factor of 30 weaker than the optimistic limits achieved with templates that describe the pseudo-data very accurately (red-dashed line in Fig.~\ref{limitplot}).  There is, however, hope for significantly more accurate and precise theoretical predictions~\cite{Pich:2013lsa, Shekhovtsova:2014hda, Bevan:2014iga}.

Because the theoretical calculation for $\tau^-\rightarrow \nu \pi^-\pi^+\pi^-$ uses input parameters measured from $\tau^-\rightarrow \nu \pi^-\pi^+\pi^-$ data, it is important to ask whether it may be biased to use such a calculation to place limits on new-physics signals.  
The resonance parameters for the $a_1^-(1260)$ can differ significantly depending on the the model and the process in which they are measured~\cite{Yao:2006px, Pich:2013lsa}.  Detailed analyses of the present and upcoming data, as well as further theoretical developments, can be expected to better model the shape of the hadronic current~\cite{Pich:2013lsa, Shekhovtsova:2014hda, Bevan:2014iga, Aushev:2010bq}.  
While this is a non-trivial issue to address, we suspect it is likely that  different physics effects can be disentangled if one goes beyond just the shape of the $m_h$ distribution (as is typically done in fits to the data) by instead analyzing the full $E_h-m_h$ phase space.  The presence of a heavy neutrino considered in this analysis manifests itself as a round endpoint structure in the $E_h-m_h$ phase space, which is quite different from the effects of resonant production and non-perturbative QCD effects.   Also, the possibility of this bias may motivate developing methods that depend minimally on the theoretical modeling of hadronic physics and instead look directly for the shape of the heavy-neutrino-endpoint signature.

\section{Discussion and Conclusion}
\label{conclusion}

We consider that the standard model (SM) is augmented by a heavy neutrino, $\nu_4$, which can couple to the tau via weak charged-current interactions, controlled by the parameter $|U_{\tau4}|^2$.  Specifically, the value of $|U_{\tau4}|^2$ is the probability that $\nu_4$ will interact with the tau, and $(1-|U_{\tau4}|^2)$ is the probability that the tau interacts with any ``light'' neutrino.  The mass of the heavy neutrino, $m_4$, can be (in principle) anywhere between the eV and GUT scales.  If $\nu_4$ is ``light,'' then the value of $|U_{\tau4}|^2$ is nearly impossible to constrain or measure, even at oscillation experiments.  On the other hand, if $\nu_4$ is heavier than $\mathcal{O}$(1 MeV), then it could decay into detectable SM particles, and the value of $|U_{\tau4}|^2$ can be experimentally investigated.

The NOMAD~\cite{Astier:2001ck} and CHARM~\cite{Orloff:2002de} experiments use high-energy protons incident on fixed-targets to potentially produce heavy neutrinos in association with taus.  Downstream, these experiments attempt to measure the decay of a heavy neutrino within the fiducial region of a detector.  Both NOMAD and CHARM place limits on the value of $|U_{\tau4}|^2$ as a function of $m_4$, assuming a model where $\nu_4$ can be produced via weak charged currents but primarily decays via weak neutral currents.  The results from NOMAD and CHARM are shown in Fig.~\ref{limitplot}.
CHARM estimates that $|U_{\tau4}|^2 < \mathcal{O}(10^{-4} - 10^{-1})$ when 20 MeV $\lesssim m_4 \lesssim $ 300 MeV.
Permitting that $\nu_4$ can decay via other forces this assumption, however, can greatly shorten the lifetime of $\nu_4$, making it more probable for it to decay well before reaching the fiducial volume of the detector and rendering it more difficult to place as strong of limits on the value of $|U_{\tau4}|^2$.

The results from the DEPHI experiment~\cite{Abreu:1996pa} at LEP assumes $\nu_4$ is a (mostly) sterile neutrino, and searches for the production and subsequent decay of $\nu_4$ in $e^+e^-\rightarrow Z \rightarrow \nu_i \nu_4$ events, where $\nu_i=\nu_1,\nu_2,\nu_3$ are the three ``light'' neutrino mass eigenstates in the SM.  DELPHI assumes $\nu_4$ can interact with electrons, muons, and taus via the weak interactions, controlled by parameters $|U_{e4}|^2$, $|U_{\mu4}|^2$, and $|U_{\tau4}|^2$, respectively.  A dataset is chosen, in principle, that would contain almost all $Z \rightarrow \nu_i \nu_4$ events, where $\nu_4$ decays via the weak interactions into ``visible'' particles within the DELPHI detector.  A limit is estimated on the combination $( |U_{e4}|^2 +|U_{\mu4}|^2 + |U_{\tau4}|^2)$, and the individual value of $|U_{\tau4}|^2$ should not exceed this limit (modulo a small kinematic factor).  If so, then $|U_{\tau4}|^2 < \mathcal{O}(10^{5} - 10^{-3})$ when 1 GeV $\lesssim m_4 \lesssim $ 60 GeV.  A portion of this limit is shown in Fig.~\ref{limitplot}.

The authors of Ref.~\cite{Helo:2011yg} use the measured branching fractions of taus, $D$'s, and $K$'s  to estimate limits on $|U_{\tau4}|^2$, requiring that $\nu_4$ decays within 10m from its production point and assuming that the tau mass and lifetime are known to infinite precision.  This result is shown by the purple line, labelled (a), in Fig.~\ref{limitplot}.  If the current measured uncertainties associated with the tau mass and lifetime are taken into account, this result could weakened by a factor of 2 or 3.   The authors of Ref.~\cite{Helo:2011yg} present another limit on $|U_{\tau4}|^2$, assuming that $\nu_4$ does not decay to ``visible'' particles within 10m, which could contribute to the measured values of $\mathcal{B}r(\tau^-\rightarrow \nu\overline{\nu}\ell^-)$, where $\ell = e, \mu$.   They estimate that $|U_{\tau4}|^2<\mathcal{O}(10^{-3})$ when $0\leq m_4\lesssim 1$ GeV.  However, this result does not require that the probability for a tau decaying to very ``light'' neutrinos or heavy neutrinos must sum to unity, nor does it incorporate the uncertainties associated with the tau mass and lifetime.  We recalculate this limit in Appendix~\ref{fixedlimits}, not ignoring these effects. In this case, we find that when $m_4\approx m_\tau$, $|U_{\tau4}|^2 \lesssim 5\times10^{-3}$, and the limit becomes extremely weak when $m_4 \lesssim$ 100 MeV.  The limit is shown by the purple line in Fig.~\ref{limitplot}, labeled (b).

These estimated limits on the value of $|U_{\tau4}|^2$ using data from NOMAD, CHARM, DELPHI, and branching fractions of taus and mesons do strongly depend on the model of $\nu_4$ decay.  In particular, all analyses assume that there are no new forces beyond the weak interactions. It is possible that permitting the existence of a new interaction or new particles could significantly change the interpretation of the data.  
If so, $\nu_4$ may prefer to decay to ``invisible'' particles, like other neutrino or new light states, which would make it experimentally challenging to detect events where the $\nu_4$ decays, even if a heavy neutrino did exist.

We present a method to place limits on $|U_{\tau4}|^2$ which does not depend on the details regarding the way the heavy neutrino decays.  We utilize $e^+e^- \rightarrow \tau^+ \tau^-$ events at a $B$ factory, requiring that one of the taus decays as $\tau^-\rightarrow \nu + h^-$, where $h^-$ is a hadronic system and $\nu$ is a mass-eigenstate neutrino.  We choose to analyze events where $h^- = \pi^- \pi^+ \pi^-$ because of the large branching ratio and excellent momentum resolution.  Information regarding the presence of a heavy neutrino is contained within the energy and momentum of $h^-$, and as the value of $m_\nu$ increases, the available energy and momentum of $h^-$, i.e., its kinematic phase space, is reduced, as shown in Fig.~\ref{psplot}.  With this method, it is possible to place strong limits on $|U_{\tau4}|^2$ even if the heavy neutrino decays to light ``invisible'' states.  This relies on the ability to inclusively select $\tau^+\tau^-$ events, where one tau decays to $\pi^+\pi^-\pi^\pm$, independent of other ``visible'' or ``invisible'' particles in the event due to the $\nu_4$ decay.

By analyzing a large number of events, one can place limits on $|U_{\tau4}|^2$, independent of any assumptions regarding how $\nu_4$ may decay.  Using MC alone, we find that, with a large dataset of $\sim10$M $\tau^-\rightarrow \nu \pi^-\pi^+\pi^-$ events at a $B$ factory, it may be possible to constrain  $|U_{\tau4}|^2 < \mathcal{O}(10^{-7} - 10^{-3})$ for 100 MeV $\lesssim m_4 \lesssim 1.2$ GeV, as shown by the red line in Fig.~\ref{limitplot}.  
The shape of the red lines in Fig.~\ref{limitplot} is due to the location and shape of the $a_1^-$ resonance, which affects the number of events near the endpoint of the phase space associated with the heavy neutrino.  Additionally, the slope of the $a_1^-$ resonance affects one's ability to extract the endpoint structure, e.g., it is easier to observe the heavy neutrino endpoint in a slowly falling region as opposed to a rapidly falling region.
The limits illustrated here are an optimistic estimation; a real data analysis of this type would be dominated by non-trivial systematic uncertainties.  If the model parameters of the $a^-_1(1260)$ resonant production of $\pi^-\pi^+\pi^-$ do not accurately describe the data, the resultant limits might be weakened.  We discuss the possibility of weaker limits in Section~\ref{analysis}.  We are optimistic that the theoretical predictions will increase in precision, and a detailed analyses using real data from, e.g., Belle, BaBar, and upcoming Belle-II~\cite{Aushev:2010bq}, will be able to successfully address these challenges.

\appendix 

\section{Limits on $|U_{\tau4}|^2$ from $\mathcal{B}r(\tau^-\rightarrow \nu \overline{\nu} \ell^- )$ }
\label{fixedlimits}

Event by event, experiments cannot distinguish between a massless or massive final-state neutrino.  Thus, if a tau decays to a heavy neutrino, it can contribute to the measurement of $\mathcal{B}r(\tau^-\rightarrow \nu \overline{\nu} \ell^- )$, assuming that the heavy neutrino itself does not decay to ``visible''  particles within the detector.  The authors of Ref.~\cite{Helo:2011yg} use the uncertainties associated with $\mathcal{B}r(\tau^-\rightarrow \nu \overline{\nu} \ell^- )$, where $\ell=e,$ $\mu$, to estimate $|U_{\tau4}|^2<\mathcal{O}(10^{-3})$ when $0\leq m_4\lesssim 1$ GeV.  However, this result does not require that the probability for a tau decaying to very ``light'' neutrinos or heavy neutrinos must sum to unity.  Ignoring this can lead to a spurious result.  Here, we perform this calculation again, not ignoring unitarity. 

Given a heavy (mostly sterile) neutrino of mass $m_4$, the predicted rate of $\tau^-\rightarrow \nu \overline{\nu}  \ell^-$ at tree-level is
\begin{equation}
\Gamma(\tau^-\rightarrow \nu \overline{\nu}  \ell^-) = \frac{G_F^2m_\tau^5}{192 \pi^3} \left[ \left(1-|U_{\tau4}|^2 \right) f\left(0, \frac{m_\ell}{m_\tau}\right) + |U_{\tau4}|^2 f\left(\frac{m_4}{m_\tau}, \frac{m_\ell}{m_\tau}\right) \right],
\end{equation}
where,
\begin{equation}
f(\alpha, \beta) \equiv  12 \displaystyle\int_{(\alpha + \beta)^2}^1 dx ~\frac{(1-x)^2}{x} \left( x - \alpha^2 - \beta^2 \right) \sqrt{ x^2 - 2x\left( \alpha^2+\beta^2 \right) + \left( \alpha^2 - \beta^2 \right)^2}.
\end{equation}
We do not consider high-order corrections in this calculation.  The measured values of $G_F$, $m_\tau$, $\tau_\tau$, and the branching ratios for leptonic tau decays are~\cite{Agashe:2014kda} 
\begin{eqnarray}
G_F &=& (1.16637 \pm 0.00001) \times 10^{-5 } \text{ GeV}^{-2}, \\
m_\tau &=& 1776.82\pm 0.16 \text{ MeV}, \\
\tau_\tau = \frac{1}{\Gamma_\text{tot}}&=& (290.6\pm 1.0)\times 10^{-15} \text{ s}, \\
\mathcal{B}r(\tau^-\rightarrow \nu \overline{\nu} e^-) &=& (17.83 \pm 0.04) \%, \\
\mathcal{B}r(\tau^-\rightarrow \nu \overline{\nu} \mu^-  ) &=& (17.41 \pm 0.04) \%. 
\end{eqnarray}
Given that $\mathcal{B}r(\tau^-\rightarrow \nu \overline{\nu}  \ell^-) $ is $\Gamma(\tau^-\rightarrow \nu \overline{\nu}  \ell^-)/\Gamma_\text{tot}$, we use these measurements as inputs for a $\chi^2$ function, including both the $\tau^-\rightarrow \nu \overline{\nu} e^-$ and $\tau^-\rightarrow \nu \overline{\nu} \mu^-$ channels.  We marginalize over the uncertainties associated with $m_\tau$ and $\tau_\tau$ via nuisance parameters (the effect of the uncertainty associated with $G_F$ is negligible).  As far as we can tell, this is not performed in Ref.~\cite{Helo:2011yg}.  For a given value of $m_4$, we vary about the minimum of this $\chi^2$ by 3.84 to estimate the 95\% CL for the value of $|U_{\tau4}|^2$.  These limits are estimated with the assumptions that the heavy neutrino decays ``invisibly,'' that $|U_{e4}|^2 = |U_{\mu4}|^2=0$, and that the data selection is not strongly affected by the presence of a heavy neutrino in the final-state. As expected, the limits are strongest when $m_\tau \leq m_4$, where $|U_{\tau4}|^2 \lesssim 5\times10^{-3}$, and become extremely weak when $m_4 \lesssim$ 100 MeV.  These results only extend up to $m_4 \lesssim \mathcal{O}$(100 GeV) so perturbativity can be maintained.  The results are shown by the purple line, labeled (b), in Fig.~\ref{limitplot}.

\begin{acknowledgments}
We thank Andr\'{e} de Gouv\^{e}a for providing useful feedback and comments.  The work of AK is sponsored in part by the DOE grant No.~DE-FG02-91ER40684 and by the Department of Energy Office of Science Graduate Fellowship Program (DOE SCGF), made possible in part by  the American Recovery and Reinvestment Act of 2009, administered by ORISE-ORAU under contract no.~DE-AC05-06OR23100.  The work of SD is supported by the DOE. 
\end{acknowledgments}

\bibliography{bib}{}

\end{document}